\documentclass[fleqn,usenatbib]{mnras}

\usepackage{newtxtext,newtxmath}

\usepackage[T1]{fontenc}
\usepackage{ae,aecompl}

\usepackage{graphicx}	
\usepackage{amsmath}	
\usepackage{amssymb}	
\usepackage{gensymb}

\defcitealias{Brouillet2005}{B05}

\title[Dense gas and star formation in M31]{Dense gas and star formation in individual Giant Molecular Clouds in M31}

\author[S. Viaene]{
S. Viaene,$^{1,2}$\thanks{E-mail: sebastien.viaene@ugent.be}
J. Forbrich,$^{1,3}$
J. Fritz$^{4}$
\\
$^{1}$Centre for Astrophysics Research, University of Hertfordshire, College Lane, Hatfield AL10 9AB, UK \\
$^{2}$Sterrenkundig Observatorium, Universiteit Gent, Krijgslaan 281, B-9000 Gent, Belgium\\
$^{3}$Harvard-Smithsonian Center for Astrophysics, 60 Garden St, Cambridge, MA 02138, USA\\
$^{4}$Instituto de Radioastronom\'{i}a y Astrof\'{i}sica (IRyA-UNAM), Antigua Carrettera a
P\'{a}tzcuaro, 8701, Morelia, Michoac\'{a}n, Mexico. \\
}

\date{Accepted XXX. Received YYY; in original form ZZZ}

\pubyear{2018}

\begin{document}
\label{firstpage}
\pagerange{\pageref{firstpage}--\pageref{lastpage}}
\maketitle

\begin{abstract} 
Studies both of entire galaxies and of local Galactic star formation indicate a dependency of a molecular cloud's star formation rate (SFR) on its dense gas mass. In external galaxies, such measurements are derived from HCN(1-0) observations, usually encompassing many Giant Molecular Clouds (GMCs) at once. The Andromeda galaxy (M31) is a unique laboratory to study the relation of the SFR and HCN emission down to GMC scales at solar-like metallicities. In this work, we correlate our composite SFR determinations with archival HCN, HCO$^+$, and CO observations, resulting in a sample of nine reasonably representative GMCs. We find that, at the scale of individual clouds, it is important to take into account both obscured and unobscured star formation to determine the SFR. When correlated against the dense-gas mass from HCN, we find that the SFR is low, in spite of these refinements. We nevertheless retrieve an SFR -- dense-gas mass correlation, confirming that these SFR tracers are still meaningful on GMC scales. The correlation improves markedly when we consider the HCN/CO ratio instead of HCN by itself. This nominally indicates a dependency of the SFR on the dense-gas {\it fraction}, in contradiction to local studies. However, we hypothesize that this partly reflects the limited dynamic range in dense-gas mass, and partly that the ratio of single-pointing HCN and CO measurements may be less prone to systematics like sidelobes. In this case, the HCN/CO ratio would importantly be a better empirical measure of the dense-gas content itself.
\end{abstract}




\section{Introduction}

In recent years, a growing interest has developed in connecting Galactic and extragalactic studies of star formation, enabled by access to similar physical scales. The main synergy lies in connecting the level of detail of local studies with a wider range of physical conditions in observations of star formation in nearby galaxies. The main observational goal in both cases is to tie the occurrence of star formation to properties of the interstellar medium. The best-known extragalactic example is the empirical Kennicutt-Schmidt (KS) relation between the surface density of star formation and the surface density of the {\it total} (atomic and molecular) gas content of entire galaxies \citep{Schmidt1959,Kennicutt1998}.

At the same time, studies of star formation at local, Galactic scales, have reached an unprecedented level of detail, and it has become possible to both test extragalactic scaling relations and formulate new hypotheses. The main advantages over extragalactic studies are 1) more direct access to fundamental parameters such as masses and star formation rates, for example by using extinction mapping and YSO counting (e.g., \citealp{Lada2010}), and 2) the opportunity to validate methodologies of extragalactic measurements locally (e.g., \citealp{Heiderman2010,Chomiuk2011}).

Based on a detailed sample of local clouds (d$<$500~pc), \citet{Lada2012} importantly argued that the KS relation does not explain star formation in the local clouds. Instead, they found that an empirical dense-gas scaling relation, where the star formation rate in a cloud depends linearly on its mass of dense gas, may be a more fundamental relation than the KS relation. Additionally, it both describes the local clouds and it connects to similar reasoning for entire galaxies by \citet{Gao2004}. In a follow-up study, \citet{Lada2013} concluded that a KS scaling relation only explains star formation within but crucially not between clouds. Cloud structure results in enhanced SFRs in regions with high column densities, but this does not necessarily imply a universal sharp (Heaviside) column density threshold of star formation. They conclude that a dense-gas scaling relation provides a far better description of star formation in the local clouds than a global KS relation. In the case of the local clouds, this formalism results in a strong correlation of a cloud's SFR with the dense gas above $A_K>0.8$~mag, as confirmed by \citet{Evans2014} and \citet{Vutisalchavakul2016}.

Other than column density mapping utilizing infrared extinction, which is only available for nearby clouds, gas of different volume densities can be observed with the help of molecular tracers. This is so far the best tool in external galaxies as demonstrated by the pioneering study of \citet{Gao2004} who found a linear relation for entire galaxies between the star formation rate, as estimated from the total infrared luminosity, and the dense gas mass, as determined from the luminosity of the HCN(1-0) transition. In a picture where the corresponding total gas mass is determined from CO observations, this scenario matches that described for the local clouds by \citet{Lada2012}. Moving toward smaller scales, this picture based on infrared and HCN observations can be extended to regions of massive star formation in our Galaxy (\citealp{Wu2005,Jackson2013}, see also \citealp{Stephens2016}). 

In the local clouds, the N$_2$H$^+$ molecule is a better tracer of dense and cold gas in comparison to HCN, as shown in particular in comparison with {\it Herschel} data \citep[see e.g.][]{Forbrich2014,Shirley2015,Pety2017,Kauffmann2017}. However, N$_2$H$^+$ is even more difficult to observe in external galaxies than HCN, and the focus in extragalactic studies so far thus has been on the latter molecule. Insights from local clouds provide promising support for such studies. In particular, the results of \citet{Lada2012} indicate that the dense-gas scaling relation based on HCN and CO from \citet{Gao2004} and \citet{Wu2005} and extinction mapping of local clouds could be put on the same scale with only minor modifications. 

However, it has so far been difficult to test the correlation of HCN, CO and the SFR in individual Giant Molecular Clouds (GMCs) of size $<$100~pc in nearby galaxies, even though extensive studies on kpc scales are becoming available \citep{Usero15,Bigiel2016}. In one of the highest-resolution studies of the relation of dense gas and star formation in an external galaxy beyond the Magellanic Clouds so far, \citet{Chen2017} study a cloud complex in M51 in CO(1-0) and HCN(1-0) as well as in other transitions. At a common resolution of $\sim$~200~pc, they compare these measurements with infrared luminosities at 70~$\mu$m as an SFR tracer, and find that their results are compatible with those of \citet{Gao2004}.
In the Local Group, \citet{Braine2017} compiled a sample of GMCs in low-metallicity galaxies, effectively obtaining a resolution of $\sim$~100~pc. They find a positive trend between both HCN and HCO$^+$ emission and the star formation rate, but argue that abundance ratios can have a strong effect on the molecular line intensities in these low-metallicity environments.

One of the most important local laboratories for star formation is the Andromeda galaxy (M31). Due to its proximity, high spatial resolution of $\sim 100$ pc can already be reached with single-dish facilities like the IRAM 30m telescope. With a considerable investment of observing time, \citet{Brouillet2005} conducted the first and so far largest survey of HCN in M31 molecular clouds, with 16 pointings, providing us with a first sample of {\it individual GMCs throughout M31} with HCN detections, which \citet{Chen2017} have already shown to be broadly compatible with the relation found by \citet{Gao2004}. However, previous extragalactic studies of HCN as a dense-gas tracer and its relation to the SFR have not only mostly focused on considerably larger (kpc) scales but also on the total infrared luminosity as a simple SFR estimate.
The latter becomes a less reliable tracer of the total star formation at the small scales in M31 (see Sect.~\ref{sec:data}).

Using the \citet{Brouillet2005} measurements, we here present a first study of how improved SFR estimates, taking into account both obscured and unobscured star formation, relate to the dense gas content of individual GMCs in an external galaxy. We adopt the same distance to M31 as \citet{Brouillet2005}; $780$ kpc.

\section{Data and sample} \label{sec:data}

\begin{figure}
	\includegraphics[width=0.5\textwidth]{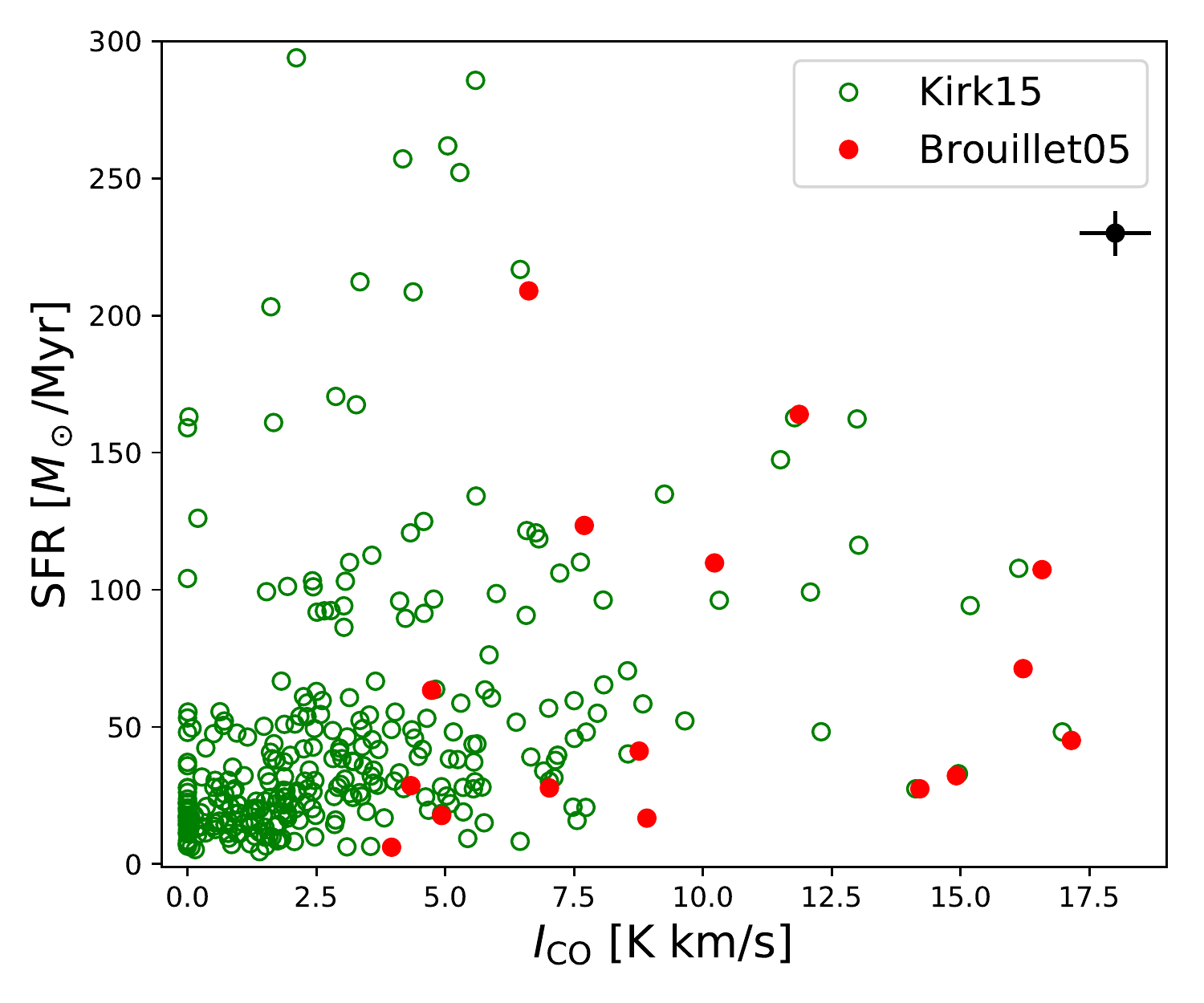}
    \caption{Star formation rate as a function of integrated CO intensity for molecular clouds in Andromeda. The green points correspond to the catalog of \citet{Kirk2015}, the red data points are the regions targeted by \citet{Brouillet2005}. The average uncertainties are shown in black in the top right corner.}
    \label{fig:CO_SFR}
\end{figure}

The starting point of our comparison is the sample of \citet{Brouillet2005}, hereafter \citetalias{Brouillet2005}, who target 16 dark clouds in the disk of M31. These clouds have been observed using the IRAM 30m telescope to detect the HCN and HCO$^+$ molecular lines. The clouds all belong to the disk of M31, with galactocentric radii between 2.4 and 15.5 kpc and are all covered in the CO(1-0) map\footnote{While \citetalias{Brouillet2005} use a preliminary version of this map, we use the published version. Excellent agreement was found between the quoted CO fluxes of \citetalias{Brouillet2005}, and our own measurements.} by \citet{Nieten2006}. For clarity, the cloud coordinates can be found in Table~\ref{tab:clouds}, converted to RA and Dec from the \citetalias{Brouillet2005} positions.

Each pointing has a FWHM of $27.5$ arcsec \citepalias{Brouillet2005}, which corresponds to a physical size of $104$ pc along the major axis in the disk of Andromeda. To accurately measure the star formation rate (SFR) in each pointing, we use the high-resolution (beamsize $\sim 6^{\prime\prime}$) SFR map created by \citet{Ford2013}. They combined the GALEX $FUV$ and {\it Spitzer}-MIPS $24 \, \mu$m maps for Andromeda, and use the {\it Spitzer}-IRAC $3.6 \, \mu$m map to correct both maps for the contribution of evolved stars. This is an important step for the SFR calculation in M31, as the contribution of evolved stars to both the $FUV$ flux and dust heating is significant \citep[see also][and references therein]{Viaene2017}. The corrected $FUV$ and $24 \, \mu$m maps are converted into SFR based on the formalism of \citet{Leroy2008}. This composite tracer is sensitive to a range of star formation timescales, as discussed in \citet{Kennicutt2012}. We refer the reader to \citet{Ford2013} for more details. 

The SFR and CO intensity of the \citetalias{Brouillet2005} clouds are plotted in Fig.~\ref{fig:CO_SFR}. For comparison, we also extract and plot these properties for all M31 molecular clouds found by \citet{Kirk2015}, based on cold dust emission. In this plot, there are two things to note. First is that the small sample of \citepalias{Brouillet2005} covers a wide range of CO and SFR values in Andromeda, with a slight bias towards high CO intensities. This is by design to increase the chances of HCN detections. Second, the correlation between CO and SFR is weak for both samples. For the molecular clouds in Andromeda, there is no strong link between the absolute content of molecular gas, and the rate of star formation. We note that these measurements are made for the same aperture size (i.e. one FWHM), so the correlation will not change when plotting surface density quantities.

\section{Dense gas in M31}

\begin{figure}
    \includegraphics[width=0.45\textwidth]{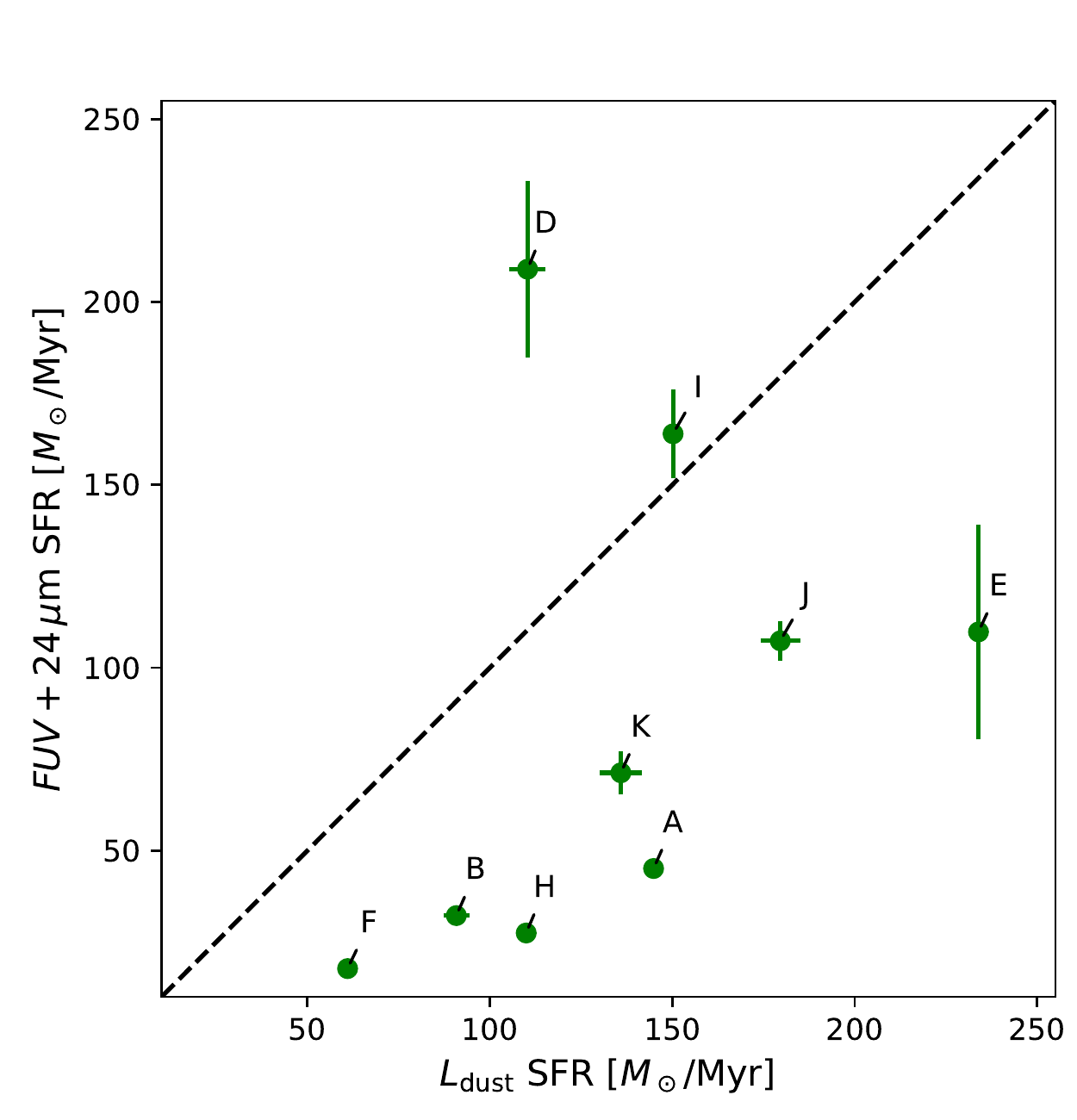} 
    \caption{Comparison of the SFR for the HCN-detected regions. The SFR on the abscissa was derived as in \citet{Gao2004}. This linearly scales with the total dust luminosity, computed from integrating the full infrared SED \citep{Viaene2014}. The SFR on the ordinate was derived from a combination of $FUV$ and $24 \, \mu$m emission, correcting for the contribution of evolved stars \citep{Ford2013}. The black dashed line indicates a 1:1 correspondence.}
    \label{fig:Ldust_SFR}
\end{figure}

\begin{figure*}
    \includegraphics[width=0.80\textwidth]{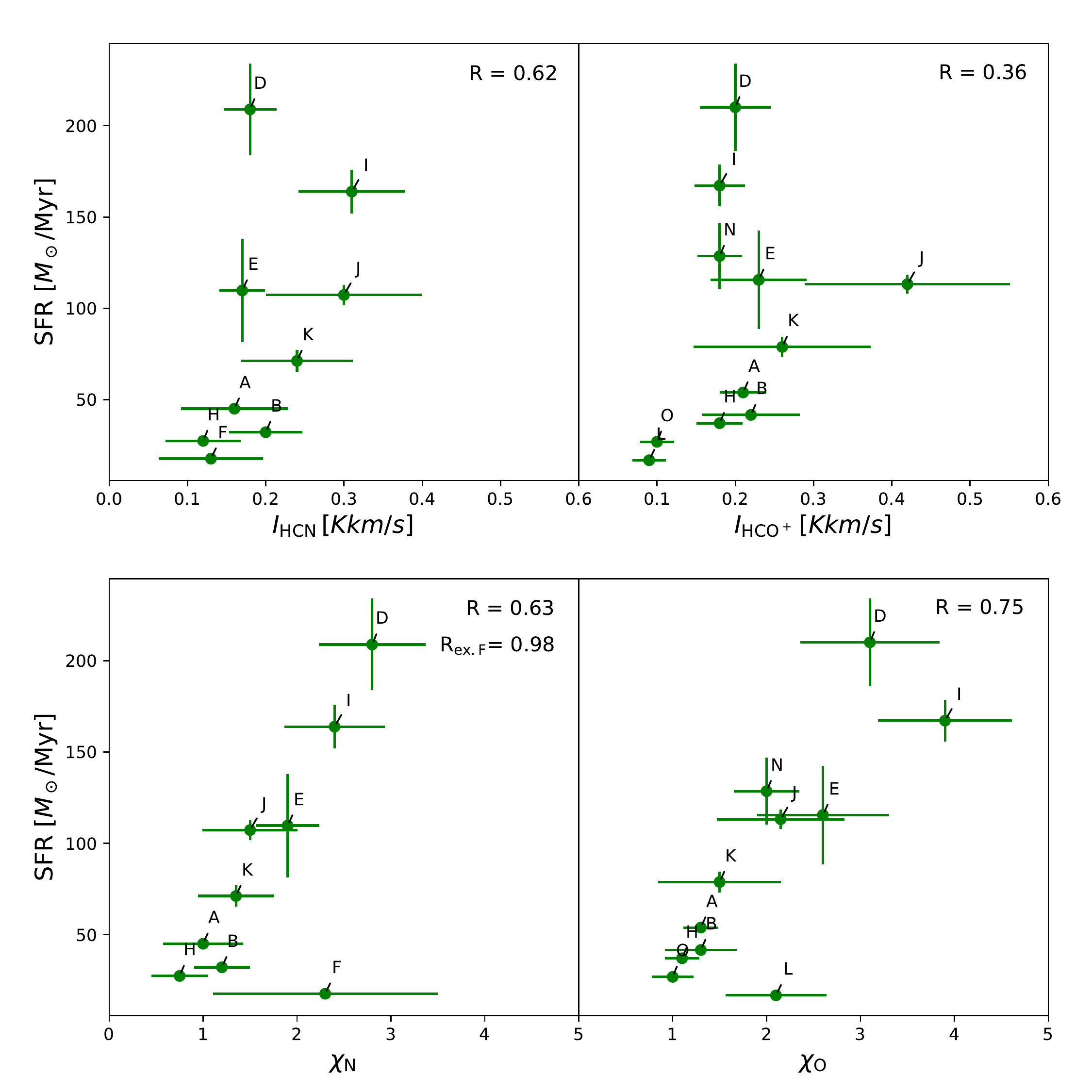} 
    \caption{Star formation and dense gas in the clouds detected by \citetalias{Brouillet2005}. Top row: Total star formation rate versus the HCN and HCO$^+$ integrated line intensities. Bottom row: Total star formation rate as a function of the HCN/CO and HCO$^+$/CO line ratios, $X_N$ and $X_O$, respectively (in percent). The Spearman rank coefficient R is shown in the top right of each panel. The errors are conservatively estimated to reflect an assumed relative positional uncertainty of 1/4 IRAM beam, plus the measurement uncertainty on the parameters themselves.}
    \label{fig:XN_SFR}
\end{figure*}

The lack of correlation between CO intensity and SFR points towards a more subtle relationship between molecular gas and star formation. While the critical density for CO(1-0) lies around $10^3$ cm$^{-3}$, this is much higher ($10^{5.7}$  cm$^{-3}$) for HCN(1-0) \citep{Shirley2015}, making the transition more sensitive to dense molecular gas.

The HCN(1-0) emission line is detected in 9 of the 16 pointings from \citetalias{Brouillet2005}. Before we go into the analysis of the dense gas properties of these clouds, we compare our new SFR estimate (from the \citealt{Ford2013} map) with what one would obtain following the classic linear conversion between SFR and the total dust luminosity (see equation 9 in \citealt{Gao2004}). While this is the preferred SFR tracer in previous HCN-SFR studies \citep[see e.g.][for an overview]{Usero15}, we do not find a clear correspondence on physical scales of 100 pc (see Fig.~\ref{fig:Ldust_SFR}). Some of the pointings, notably clouds D and I, exhibit clear signs of unobscured star formation (see Fig.~\ref{fig:SDSScutouts}). This is not picked up in $L_\mathrm{dust}$, but is incorporated in the $FUV + 24\mu$m tracer. On larger scales, one can expect the fraction of obscured SFR to dominate, especially in the dust-rich star-forming galaxies which were investigated so far.

In addition, $L_\mathrm{dust}$-based SFRs generally yield higher rates of star formation. This is particularly true for M31, where the dust is in fact mainly heated ($60-90$ \% for these regions) by evolved stars \citep{Viaene2017} thus artificially increasing this SFR estimate. In the remainder of this work, we therefore rely on the composite SFR tracer, which includes unobscured star formation, and where the contribution of the evolved stars has been removed.

With our more comprehensive SFR estimate we confirm a general trend between HCN line emission and the SFR (see the top left panel in Fig.~\ref{fig:XN_SFR}), with a Spearman correlation coefficient $R$ of $0.62$. For completeness, we also include the HCO$^+$ line intensities in Fig.~\ref{fig:XN_SFR} (top right panel). There is still a positive correlation, but the scatter is larger ($R = 0.36$).

We find an even better correlation when plotting the SFR against the HCN/CO ratio (Fig.~\ref{fig:XN_SFR}, lower left panel). We find a correlation coefficient of $R = 0.63$, drastically improving to $R = 0.98$ when removing the marginal outlier point F (see below). Both values only leave a small chance that this correlation is a statistical coincidence. 
A linear fit to the data, taking into account their uncertainties, yields
\begin{equation}
SFR = (88 \pm 11) X_N - (50 \pm 16),
\end{equation}
which yields SFR in $M_\odot / \mathrm{Myr}$ from $X_N = 100 I_\mathrm{HCN}/I_\mathrm{CO}$. 

The trend between SFR and the HCO$^+$/CO line ratio also improves when compared to the SFR vs. HCO$^+$ relation, but the scatter is slightly larger ($R = 0.75$). \citet{Braine2017} recently noted that these line ratios are indeed positively correlated with the $24 \, \mu$m flux, a proxy for obscured star formation. Their sample was a set of GMCs in low-metallicity Local Group galaxies at a comparable physical resolution $\sim 100$ pc. Fig.~\ref{fig:XN_SFR} presents, to our knowledge, the first comparison for extragalactic GMCs at $\sim 100$ pc scales in a normal-metallicity environment. In the remainder of the paper, we explore the strongest relationship (SFR vs. $X_N$) in more depth.

First, we perform a rigorous error analysis. The lower left panel in Fig.~\ref{fig:XN_SFR} shows a notable outlier: cloud F, the pointing with the lowest SFR and only a marginal, low-S/N HCN detection. In the uncertainty on each SFR measurement, we include the basic uncertainty in the SFR map (taken from \citealt{Ford2013}), plus a positional uncertainty -- mainly since we do not know the relative positional accuracy of the various maps used here. We perform a bootstrapping analysis, creating 10000 new pointings drawn from a normal probability distribution centered a the original pointing, with a conservatively\footnote{\citet{Nieten2006} report a pointing uncertainty of $<5^{\prime\prime}$.} chosen spread of $6.9^{\prime\prime}$ (1/4 times the IRAM FWHM beam). This yields an average pointing uncertainty of $8 \%$ in SFR, which is the dominant error. 

Analogously, we find an average pointing uncertainty of $5 \%$ from the CO map. We combine this with the HCN detection uncertainties quoted in \citetalias{Brouillet2005} for the uncertainty on $X_N$. Note that the uncertainties quoted on the HCN emission are already conservative estimates based on the information provided in \citetalias{Brouillet2005}. The uncertainties on $X_N$ are comparable with the scatter in the $X_N\sim$SFR relationship, which is reassuring. The only exception is indeed cloud F, which turns out to have a large uncertainty ($51 \%$) on the HCN detection. \citetalias{Brouillet2005} already noted that this cloud may be quiescent, and treat it as an outlier in their further analysis. Given the high uncertainty enshrouding the HCN detection in cloud F, we do not speculate any further on the nature of this outlier. We note that the Spearman coefficient becomes $R = 0.98$ when removing point F, and the $p$-value becomes  $3.3\times 10^{-5}$, making the correlation much more significant.

\begin{figure}
    \includegraphics[width=0.45\textwidth]{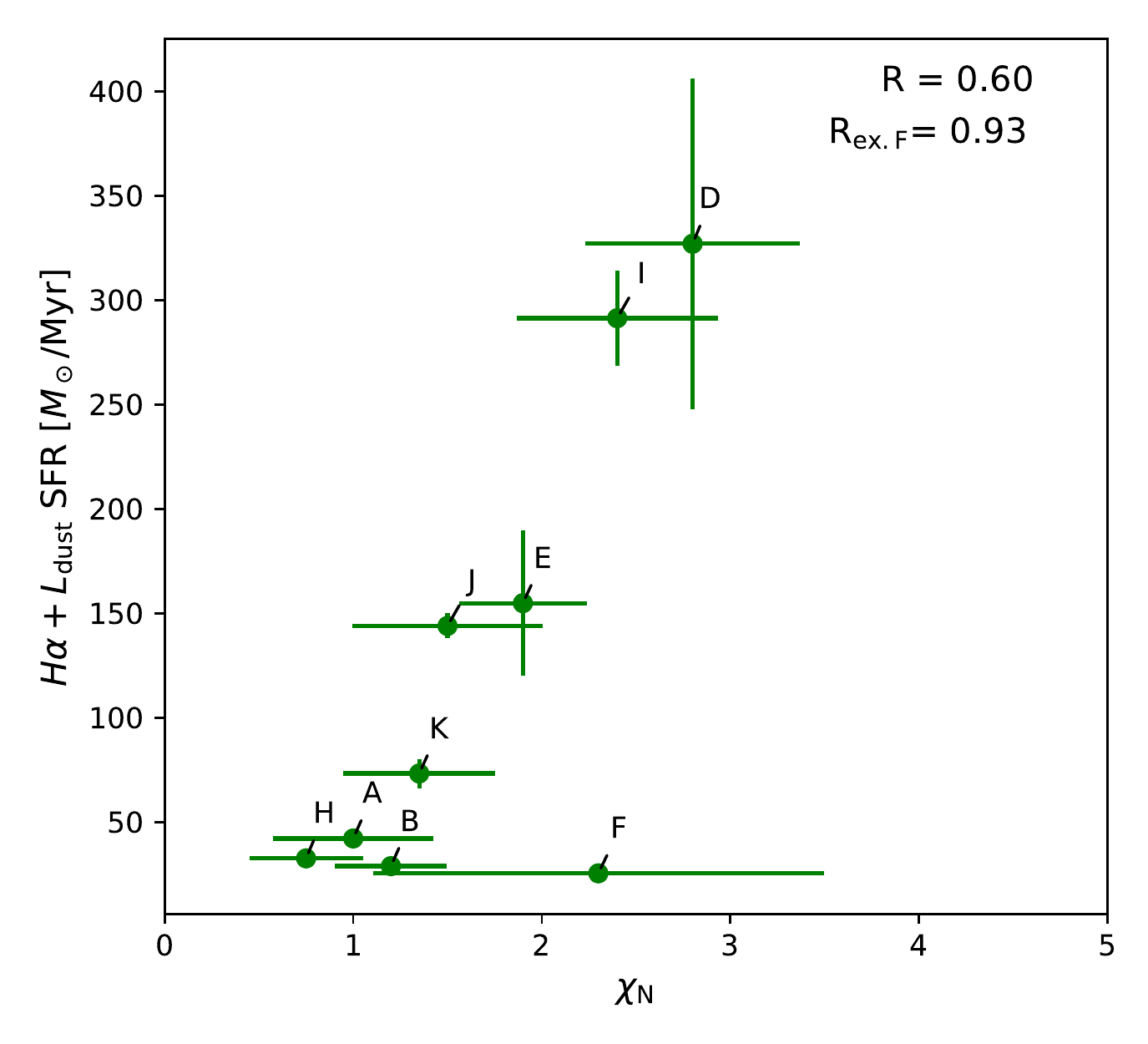} 
    \caption{Star formation rate as a function of the HCN/CO line ratio, $X_N$. This SFR estimated is based on $H\alpha + L_\mathrm{dust}$ emission. The errors are conservatively estimated to reflect an assumed relative positional uncertainty of 1/4 IRAM beam, plus the uncertainty on the SFR and $X_N$ itself.}
    \label{fig:XN_HaLdust}
\end{figure}

Second, we test this correlation using an independent measure of the SFR. We chose to use a combination of H$\alpha$ luminosity and total dust luminosity, following the prescription of \citet{Kennicutt2012}. The $H\alpha$ map from \citet{Devereux1994}) was used, combined with the map of total dust luminosity from \citet{Viaene2014}. The $X_N$ vs. SFR correlation for this particular calibration is shown in the bottom panel of Fig.~\ref{fig:XN_HaLdust}. The same trend arises (including outlier F), although shifted to slightly higher SFR. We suspect this is due to evolved stars contributing to the dust heating in M31 \citep[see][]{Viaene2017}, which is not subtracted in the $L_{H\alpha}+L_\mathrm{dust}$ calibration, but is corrected for in the $FUV + 24\, \mu$m calibration. It is not the goal of this paper to compare SFR tracers, so we will continue to use the composite $FUV + 24\, \mu$m tracer in the remainder of this work. 

\begin{figure}
    \includegraphics[width=0.45\textwidth]{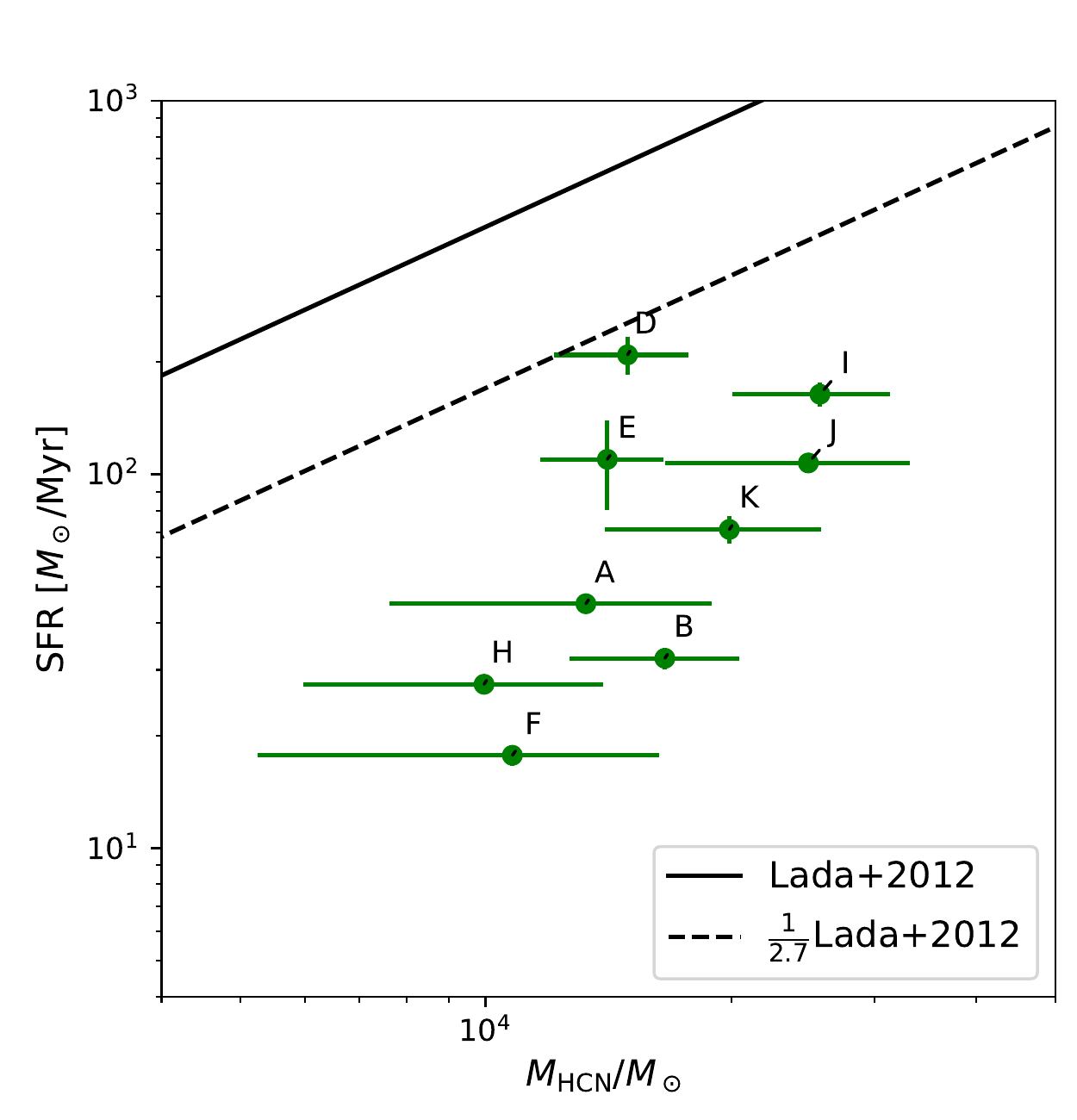} 
    \caption{Star formation rate as a function of dense gas mass as traced by HCN(1-0) ($M_\mathrm{HCN}$). The solid black line correspond to the \citet{Lada2012} relation for local Milky Way clouds. The dashed line is that same relation, but rescaled to correspond to the \citet{Gao2004} data.}
    \label{fig:HCN_SFR}
\end{figure}

The HCN line intensities can be converted to a dense gas mass. We assume $\alpha_\mathrm{HCN} = 10 \, M_\odot$(K~km/s pc$^2$)$^{-1}$ to remain consistent with \citet{Gao2004}. This yields $M_\mathrm{HCN}$ of $1-3 \times 10^4 M_\odot$. Fig.~\ref{fig:HCN_SFR} shows their relation to the SFR, together with the SFR-dense gas mass relation found by \citet{Lada2012} for local Milky Way clouds. Their SFR is based on counting young stellar objects and they find a calibration offset of 2.7 with respect to the \citet{Gao2004} relation. We therefore also show the \citet{Lada2012} relation, divided by 2.7.

While a weak trend exists between the dense gas mass and the SFR, the data does not match either of the above two relations. This was already noted by \citet{Chen2015}, who compared the \citetalias{Brouillet2005} regions in the $L_\mathrm{IR}$-$L_\mathrm{HCN}$ parameter space to sub-kpc regions in M51 and global galaxy measurements. They find that the M31 regions lie below the main trend, which is also reflected in Fig.~\ref{fig:HCN_SFR}. The average $L_\mathrm{IR}/L_\mathrm{HCN}$ ratio is $433 \pm 175 \, L_\odot$(Kkm/s pc$^2$)$^{-1}$ for the M31 regions. This lies well below the \citet{Gao2004} ratio ($900 \, L_\odot$(Kkm/s pc$^2$)$^{-1}$), but in between the values for central regions in M51 ($388 \pm 47 \, L_\odot$(Kkm/s pc$^2$)$^{-1}$) and regions in the outer disk of M51 ($691 \pm 156\, L_\odot$(Kkm/s pc$^2$)$^{-1}$)

In addition to the intrinsic offset to the \citet{Gao2004} relation, Fig.~\ref{fig:Ldust_SFR} also reveals an important offset between $L_\mathrm{IR}$-based SFR and our composite tracer. The combination of these two effects lead to rather high nominal dense-gas depletion times for the M31 clouds, which lie between 70 and and 600 Myr, although the uncertainties in the latter are substantial, driven by the uncertainties in the HCN masses.

\section{Discussion}

\subsection{GMC properties}

Before moving on to interpret their behaviour, it is important to verify the nature of the structures under investigation, and reflect on our methods to estimate their properties.

First, we verify the class of the probed regions. \citet{Brouillet2005} classify them as GMCs, based on the CO map, which has limited resolution ($23^{\prime \prime}$ beam). We reassess their classification by inspecting composite $gri$ SDSS images\footnote{\url{http://www.sciserver.org}} of the target areas (see Fig.~\ref{fig:SDSScutouts}). The optical colour maps provide a better resolution ($1-2^{\prime \prime}$ seeing) than the CO map, and the denser ISM can be traced efficiently through reddening by dust. The HCN-detected clouds can all be associated with continuous structures of dust extinction. These structures are several 100 pc in size, with multiple star-forming regions, suggesting they are giant molecular associations (GMAs). The HCN pointings pick up entities ($\sim100$ pc in diameter) within these structures. This confirms the classification of \citet{Brouillet2005} in that we are most likely tracing individual GMCs.

Second, it is difficult to assess whether we are observing the same physical structure in CO, HCN and in the SFR map. The latter has a resolution of $6^{\prime\prime}$, so we summed pixels within a circular aperture of diameter $27.5^{\prime\prime}$. As discussed above, we then allow for the possibility of offsets between the SFR, CO, and HCN maps when cross-matching the data. This ensures that any significant source structure that falls just outside of the nominal beam at least gets included in the error budget.

Third, the inclined view of the M31 disk ($i \sim 77 \degree$) can cause line-of-sight effects. Two of the pointings (A and E) have two distinct velocity components, suggesting there are at least two different structures in the beam along the line-of-sight. In both cases, we used the ones with the highest HCN intensity (which corresponds to the highest CO intensity as well). However, for the SFR estimate, we have no velocity information and are in essence summing up the combined signal along the line-of-sight. However, even without pointings A and E, the correlation would persist. Nevertheless, it is possible that there are more unknown offsets between the origin of HCN, CO and SFR within the 100 pc beam, or along the line-of-sight. Future interferometric studies could resolve such offsets.

\subsection{Star formation rates}

We find unusually low SFRs for the M31 GMCs. They are a factor of 2--3 lower than those from $L_{\rm dust}$, which in turn are a factor of 2.7 lower than SFRs from YSO counting \citep{Chomiuk2011, Lada2012}, resulting in SFRs that are almost an order of magnitude lower than calibrations suggested by YSO counting. This in turn leads to unusually long and unrealistic depletion times of up to 600~Myr. The difference could point to stochastic sampling of star formation in these regions. If the SFR calibration assumes a fully sampled IMF and instead individual regions may have less overall emission from massive stars, then the resulting SFRs will be biased toward lower numbers.

Alternatively, individual clouds could have star formation timescales different from those usually assumed for star-formation recipes calibrated for extragalactic studies. However, \citet{Kennicutt2012} report a mean age range of 3, 5, 5 and 10 Myr for H$\alpha$, $24 \,\mu$m, TIR and FUV, respectively. These are still relatively short timescales, which are realistic even for individual clouds where stars are formed on a short timescale. SFR scaling relations are expected to break down at small spatial scales \citep[see e.g.]{Kennicutt2012,Kruijssen2014}. The SFR map of \citet{Ford2013} has a resolution of 6 arcsec, but we sum all pixels within a circle of 27.5 arcsec diameter. The SFR calibration is linear, which means our effective resolution is still 100 pc. 

Scaling relations involving SFR were found to be still valid at 140 pc scales in M31 \citep{Viaene2014}, albeit with some scatter. Also, despite the above concerns, we still observe a tight correlation between SFR and $X_N$. This indicates that, if not the absolute calibration, at least the relative differences in SFR are genuine, even for GMC-sized structures. A detailed study of the role of physical scales is beyond the scope of this paper since it will have to be addressed with more comprehensive dense-gas and SFR data.

\subsection{SFR and the HCN/CO ratio}

Our results suggest a strong connection between the star formation rate and the HCN/CO line ratio. This ratio is often interpreted as a dense gas fraction since it was thus proposed by \citet{Gao2004}. These authors find constant HCN/CO ratios for star-forming galaxies, but a higher ratio for starburst galaxies. In the local Milky Way clouds studied by \citet{Lada2012}, a clear linear link between the absolute dense gas mass (derived from dust extinction) and the SFR excludes a tight correlation between the SFR and the HCN/CO ratio. These two studies both cover a dynamic range of two orders of magnitude in dense gas mass. Due to sensitivity limitations, our M31 sample is limited to a dynamic range in the dense gas mass of a factor of about only 2.4 and by a factor of about 7.5 in the total gas mass. Despite this limitation, which allows only for a tentative search for trends, clear correlations still emerge, and using the HCN/CO ratio leverages the larger dynamic range of the CO measurements.

The reliability of HCN as a tracer of dense gas is matter of ongoing debate \citep[see e.g.][and references therein]{Costagliola2011, Pety2017, Schap2017, Barnes2017, Kauffmann2017}. Collisional excitation produces HCN(1-0) emission above a critical density, and our understanding of collisional excitation continues to improve \citep{HernandezVera2017}, but depending on the environmental conditions, observable emission may occur at densities significantly below the critical density (e.g., \citealp{Goldsmith2017}). Infrared pumping of HCN is also possible \citep[see e.g.][]{Imanishi2017}, but only when a very bright MIR source (e.g., an AGN) is present. \citet{Gao2004} did not find evidence for IR pumping in their sample, and even in extreme cases, the contribution to the line emission remains below $20$ \% \citep{Vollmer2017}. Additionally, M31 has a well-known metallicity gradient \citep{Mattsson2014}, which can change the amount of nitrogen in the gas phase and thus change HCN abundances. Nonetheless, one can argue that, globally, the physical conditions of the regions in this work are comparable since all of them are part of the disk of M31.

The apparent link between SFR and the HCN/CO ratio does not necessarily imply a physical contradiction with the locally observed correlation between the SFR and the dense-gas mass. Our hypothesis would instead be that when considering single-pointing measurements on cloud scales, the ratio of HCN and CO could be the empirically cleanest measure of the dense gas content. Determining the dense-gas mass from a single pointing may be impacted by systematic effects from additional parameters, like beam sidelobes, hence the poorer correlation in Fig.~\ref{fig:HCN_SFR}. 

Deriving $M_\mathrm{HCN}$ requires the explicit assumption of a size scale (the FWHM main beam) where all emission is thought to be contained. Since there is only one pointing per cloud, we do not know a priori how well-justified this assumption is. If, for example, the molecular peak or any other significant emission is outside of the FWHM beam, the ratio of HCN/CO, both affected in similar ways from emission caught in the outer sensitivity pattern, may be a better empirical measure of the dense-gas content of the inner cloud than just HCN, assumed to originate from within the FWHM beam and thus in tendency under-estimated if the peak is outside of the main beam. 

It is clear that the observed correlation raises several intriguing questions, and can reveal unknown aspects of star formation on cloud scales. However, the limited sample size largely prevents us from drawing firm conclusions. A larger sample of molecular clouds and spatially resolved maps are needed to confirm the exact nature of the correlation.

\section{Summary and conclusions}

In our comparison of the dense gas content and star formation rates of individual GMCs at the scale of 100 pc in M31, we find the following:

\begin{itemize}
\item To inspect the target regions further, at the highest available resolution, we study them qualitatively in optical dust extinction. We conclude that the HCN pointings contain small fractions of GMAs, and most likely individual GMCs.
\item At these scales, a composite (obscured + unobscured) SFR tracer is preferred over the total infrared luminosity as several clouds show clear evidence of unobscured star formation. Also, this tracer importantly corrects for significant dust heating by evolved stellar populations.
\item This sample of GMCs in M31 is covering a representative dynamic range in both SFR and CO luminosity when compared to analogous measurements for a comprehensive cloud catalog.
\item For the nine GMCs with an HCN detection, the SFR correlates with the HCN and HCO$^+$ luminosities and thus dense gas mass. This correlation is broadly consistent with previous extragalactic and Galactic results.
\item The correlation with SFR improves considerably when instead of the HCN luminosity we consider the ratio of the clouds' HCN and CO luminosities. A similar but weaker effect is seen when considering the HCO$^+$ emission.
\item This correlation suggests that the SFR estimates are useful, even on these small scales, but not necessarily in an absolute sense since the resulting nominal depletion times are too long when compared to individual local clouds.
\end{itemize}

In the picture originally proposed by \citet{Gao2004}, where the total mass of a cloud is measured in CO and the dense-gas mass is measured in HCN, our finding thus would suggest a correlation of the SFR with a nominal dense gas \textit{fraction} on cloud scales, in contradiction to local results. However, we hypothesize that the reason for this improved correlation is partly due to a) the limited dynamic range in the dense gas mass, such that the HCN/CO ratio leverages the larger dynamic range in total gas mass from CO measurements, and b) to systematic uncertainties that are shared by both the HCN and CO measurements, such that they cancel out in the ratio. Importantly, the HCN/CO line ratio would then correspond to a measure of the dense gas mass for this specific case of individual single-dish measurements.

\section*{Acknowledgements}
The authors wish to thank Matthew Smith, who kindly shared the high-resolution SFR map of M31, and Charles Lada, Sebastien Muller, and Laurent Loinard for their valuable feedback and discussions.
S.V. acknowledges the support of the BOF of Ghent University, and of the FWO mobility fund.




\bibliographystyle{mnras}
\bibliography{references} 



\appendix

\section{Cloud sample}

\begin{figure*}
	\includegraphics[width=\textwidth]{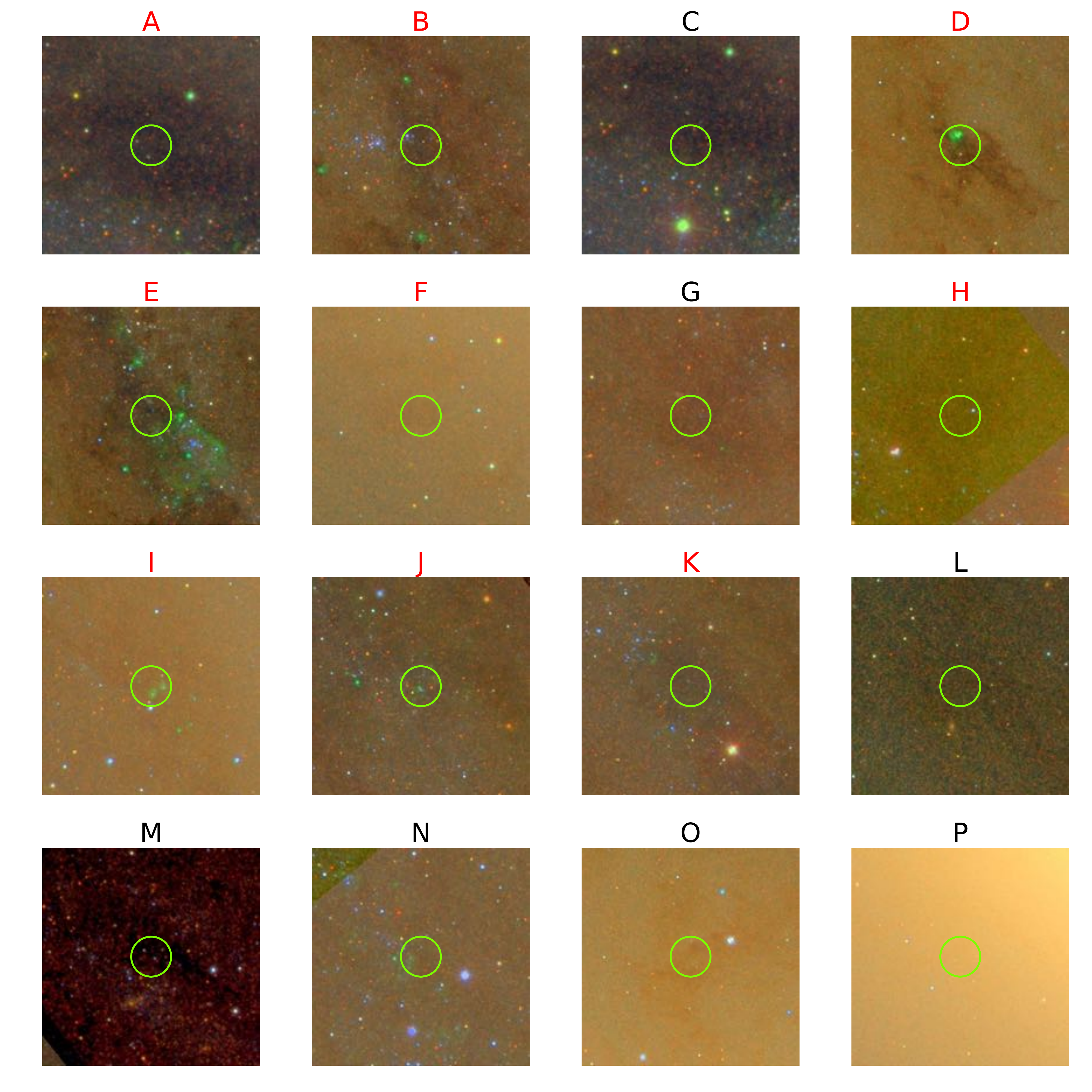}
    \caption{Composite SDSS $g$ (blue) $r$ (green) $i$ (red) colour images for the regions targeted by \citet{Brouillet2005}. The green circle indicates the pointing and beam FWHM (27.5 arcsec). HCN-detected pointings are marked with a red label.}
    \label{fig:SDSScutouts}
\end{figure*}

The coordinates for the clouds in the sample of \citet{Brouillet2005} are listed in Table~\ref{tab:clouds}, as are their SFRs. Composite SDSS colour maps are shown in Fig.~\ref{fig:SDSScutouts}.

\begin{table} 
\caption{Coordinates, star formation rates and relative uncertainties for the \citet{Brouillet2005} pointings. SFR is based on the FUV+24 micron map of \citet{Ford2013}. }
\label{tab:clouds}
\centering     
\begin{tabular}{lcccc}
\hline
Pointing & RA & DEC  & SFR  & $\sigma^\mathrm{rel}_\mathrm{SFR}$ \\
& J2000 & J2000 & $[M_\odot \mathrm{Myr}^{-1}]$ & \\
\hline
A & 00:41:05.02 & +40:38:01.7 & 45.1  & 0.04 \\
B & 00:41:13.91 & +41:08:37.9 & 32.2  & 0.09 \\
C & 00:41:04.99 & +40:37:31.7 & 41.2  & 0.05 \\
D & 00:41:30.14 & +41:04:53.7 & 209.0 & 0.20 \\
E & 00:41:00.45 & +41:03:31.1 & 109.8 & 0.42 \\
F & 00:42:41.13 & +41:07:19.9 & 17.7  & 0.09 \\
G & 00:44:26.88 & +41:38:01.1 & 27.7  & 0.11 \\
H & 00:44:13.09 & +41:35:12.6 & 27.4  & 0.03 \\
I & 00:43:56.69 & +41:26:34.1 & 163.9 & 0.13 \\
J & 00:42:38.57 & +41:31:51.7 & 107.3 & 0.08 \\
K & 00:42:31.99 & +41:29:46.8 & 71.2  & 0.12 \\
L & 00:44:02.85 & +41:42:40.2 & 06.1  & 0.05 \\
M & 00:43:29.35 & +41:48:48.4 & 63.4  & 0.11 \\
N & 00:44:43.38 & +41:27:51.1 & 123.4 & 0.21 \\
O & 00:43:06.53 & +41:24:15.4 & 16.8  & 0.12 \\
P & 00:42:56.73 & +41:14:32.8 & 28.6  & 0.07 \\
\hline
\end{tabular}
\end{table}

\bsp	
\label{lastpage}
\end{document}